\documentclass[twocolumn,aps,prl,preprintnumbers,showpacs,nofootinbib]{revtex4}
\usepackage[utf8]{inputenc}
\usepackage[english]{babel}
\usepackage{amsmath}
\usepackage{amsfonts}
\usepackage{amssymb}
\usepackage{epsfig}
\usepackage{graphics,psfrag,rotating}
\usepackage{graphicx}
\usepackage{dcolumn}
\usepackage{bm}
\usepackage{epstopdf}
\usepackage{color}
\usepackage[usenames,dvipsnames,svgnames]{xcolor}
\usepackage[colorlinks=true,
            linkcolor=red,
            urlcolor=gray,
            citecolor=blue]{hyperref}
  \usepackage{hyperref}

\newcommand{\lsim}   {\mathrel{\mathop{\kern 0pt \rlap
{\raise.2ex\hbox{$<$}}}
 \lower.9ex\hbox{\kern-.190em $\sim$}}}
\newcommand{\gsim}   {\mathrel{\mathop{\kern 0pt \rlap
{\raise.2ex\hbox{$>$}}}
\lower.9ex\hbox{\kern-.190em $\sim$}}}

\def\3nab{\tilde{\nabla}}

\def\tl{\tilde}
\def\hsp5{\hspace{5mm}}

\def\case#1/#2{\textstyle\frac{#1}{#2}}

\def\ber {\begin{eqnarray}}
\def\eer {\end{eqnarray}}
\def\bea {\begin{eqnarray}}
\def\eea {\end{eqnarray}}

\def\bc {\begin{center}}
\def\ec {\end{center}}
\def\case#1/#2{\frac{#1}{#2}}

\newcommand{\bw}{\begin{widetext}}
\newcommand{\ew}{\end{widetext}}
\newcommand{\nn}{\nonumber\\}

\newcommand{\be}{\begin{equation}}
\newcommand{\bse}{\begin{subequation}}
\newcommand{\ese}{\end{subequation}}
\newcommand{\ee}{\end{equation}}
\newcommand{\eei}{\end{eqnarray}\indent\indent}
\newcommand{\ba}{\begin{array}}
\newcommand{\ea}{\end{array}}
\newcommand{\bal}{\begin{eqnarray}}
\newcommand{\eal}{\end{eqnarray}}

\def\case#1/#2{\textstyle\frac{#1}{#2} }
\newcommand{\nb}{\nabla}

\newcommand{\gd}{g_{ab}}
\newcommand{\bver}{\begin{verbatim}}
\newcommand{\ever}{\end{verbatim}}

\begin{document}
\title{On tidal forces in  $f(R)$ theories of gravity}
\author{
\'Alvaro de la Cruz-Dombriz$^{1,2,3,\footnote{E-mail: alvaro.delacruzdombriz [at] uct.ac.za} 
}$, 
Peter K. S. Dunsby$^{1,2,4,\footnote{E-mail:  peter.dunsby [at] uct.ac.za} }$, Vinicius C. Busti$^{1,2,\footnote{E-mail: vinicius.busti [at] iag.usp.br} }$, 
Sulona Kandhai$^{1,2,\footnote{E-mail:  kndsul001 [at] myuct.ac.za}}$
}
\affiliation{$^{1}$ Astrophysics, Cosmology and Gravity Centre (ACGC), University of Cape Town, Rondebosch 7701, Cape Town, South Africa}
\affiliation{$^{2}$  Department of Mathematics and Applied Mathematics, University of Cape Town, Rondebosch 7701, Cape Town, South Africa}
\affiliation{$^{3}$  Instituto de Ciencias del Espacio (ICE/CSIC) and Institut d'Estudis Espacials de Catalunya (IEEC), Campus UAB, Facultat de Ci\`{e}ncies, Torre C5-Par-2a, 08193 Bellaterra (Barcelona) Spain}
\affiliation{$^{4}$  South African Astronomical Observatory,  Observatory 7925, Cape Town, South Africa.}
\date{\today}
\begin{abstract}
Despite the extraordinary attention that modified gravity theories  have attracted
over the past decade, the geodesic deviation equation in this context
has not received proper formulation thus far. This equation provides an elegant way to investigate the 
timelike, null and spacelike structure of spacetime geometries. In this investigation we provide the full derivation 
of this equation in situations where General Relativity has been extended in Robertson-Walker background spacetimes. 
We find that for null geodesics the contribution arising from the geometrical new terms is in general non-zero.  
Finally we apply the results to a well known class of $f(R)$ theories,  compare the results with General Relativity predictions 
and obtain the equivalent area distance relation. 
\end{abstract}
 \maketitle
\section{Introduction}

The limitations faced by the cosmological concordance model or $\Lambda$CDM model have led cosmologists to propose a range of alternative theories.  Modifications inside the framework of General Relativity (GR), with the presence of a new component called dark energy have been proposed  \cite{de_eq}, where a possible time evolution in its energy density is encoded in the equation of state.  Another possibility 
consists of replacing the theory of gravity on large scales, where a different gravitational action may explain the current accelerated phase experienced by the Universe. In this way, instead of a new fluid driving the acceleration, this effect results directly from the geometric part of the gravitational field equations.

There are several ways of modifying the gravitational action ({\it c.f.} \cite{clifton} for a thorough review), giving rise to different modified gravity theories. One of the simplest forms is to consider functions of the Ricci scalar $R$, dubbed $f(R)$ theories \cite{rev_fR} and this class will be the focus of our investigations. These theories are constrained by a number of requirements, which include: $a)$ the positivity of the 
effective gravitational constant \cite{Pogosian}; $b)$ the existence  of a stable gravitational stage related to the presence of a positive mass for the associated scalar mode \cite{fRcha} and, last but not least, $c)$ the recovery of the GR behavior on small scales and at early times in the history of the universe in order to be consistent with Big Bang Nucleosynthesis and Cosmic Microwave Background (CMB) constraints. 
There also exist several constraints for the value of $|{\rm d}f/{\rm d}R|_{R=R_0}$, where $R_0$ holds either for the current or past cosmological background curvature. The latter constraint arises from the Integrated Sachs-Wolfe effect and correlations with foreground galaxies ({\it c.f.} \cite{Referee})\footnote{These constraints are obtained using several assumptions and are therefore in general model dependent.}. 

With the aim of providing a satisfactory explanation for a range of cosmological and astrophysical phenomenon, modified gravity theories have been studied from different points of view including the growth of density \cite{Perturbations} and gravitational waves \cite{Tensor-perturbations} perturbations,  determining the existence of GR-predicted astrophysical objects such as black holes \cite{BH} as well as research on their stability \cite{Dombriz-Saez}.

One important aspect which has not received a proper treatment so far relates to the timelike, null and spacelike structure of spacetimes in the framework of fourth order gravity theories in general and the example of $f(R)$ theories in particular. An elegant way to study this can be done through an analysis of the {\em Geodesic Deviation Equation} (GDE), also known as the {\em Jacobi equation}. This equation encapsulates many results of standard cosmology \cite{Ellis-Elst} such as the observer area distance, first derived by Mattig \cite{mat58} for the dust case, the dynamics governed by the Raychaudhuri equation \cite{Ray} and how perturbations affect the kinematics of null geodesics, leading to gravitational-lensing effects \cite{lensing}.

As a first application of the GDE in metric $f(R)$ theories, we restrict our attention to Friedmann-Lema\^{i}tre-Robertson-Walker (FLRW) spacetimes and derive the GDE for the spacelike, timelike and null geodesics. Although some attention has been paid to this equation in the Palatini formalism \cite{Iranians} and in arbitrary curvature-matter coupling scenarios \cite{Harko-Lobo}, the present investigation represents the first attempt to address this issue in the metric formalism. 

We also derive the area distance relation and study the results for a specific parameterization of $f(R)$ dark energy theories - the so-called Hu-Sawicki (HS) models \cite{Sawicki}, which provide a viable cosmological evolution and have been investigated in a range of astrophysical and cosmological situations.

This paper is organized as follows. In Section \ref{Section2} we briefly review the $1+3$ decomposition of variables which allows us to obtain the cosmological equations for $f(R)$ theories 
in the metric formalism, assuming a FLRW background.  Section \ref{Section3} is devoted to studying the GDE in the context of $f(R)$ theories for homogeneous and isotropic backgrounds and a derivation of the observer area distance formula is presented in Section \ref{Section4}.  In order to characterize the background FLRW cosmology,  a dynamical system approach is described in Section \ref{Section5} and this is used in Section \ref{Section6} to illustrate the rich phenomenology of the evolution of the GDE for 
the HS class of $f(R)$ models. This demonstrates how the standard GR geodesic deviation evolution is distorted when new non-constant terms are included in the gravitational action.  Finally, in Section \ref{Section7} we present the conclusions and give an outline of future work to be done.

\section{The Cosmological Equations for $f(R)$  Gravity} \label{Section2}
\subsection{Basic Notation}
Unless otherwise specified,  natural units ($\hbar=c=k_{B}=8\pi G=1$) will be used throughout this paper. Latin indices run from 0 to 3,
the symbol $\nabla$ represents the usual covariant derivative, we use the
$(-,+,+,+)$ signature and the Riemann tensor is defined by
\begin{eqnarray}
R^{a}{}_{bcd}=W^a{}_{bd,c}-W^a{}_{bc,d}+ W^e{}_{bd}W^a{}_{ce}-
W^f{}_{bc}W^a{}_{df}\;,
\end{eqnarray}
where the $W^a{}_{bd}$ are the Christoffel symbols (i.e., symmetric in
the lower indices), defined by
\begin{equation}
W^a{}_{bd}=\frac{1}{2}g^{ae}
\left(g_{be,d}+g_{ed,b}-g_{bd,e}\right)\;.
\end{equation}
The Ricci tensor is obtained by contracting the {\em first} and the
{\em third} indices
\begin{equation}\label{Ricci}
R_{ab}=g^{cd}R_{cadb}\;.
\end{equation}
Finally, the action for $f(R)$-gravity can be written in these units as
\begin{equation}
\label{lagr f(R)}
\mathcal{A}=\int {\rm d}^4 x \sqrt{-g}\left[\frac12 f(R)+{\cal L}_{m}\right]\;,
\end{equation}
where $R$ is the Ricci scalar, $f$ is a general differentiable 
(at least $C^2$) function of the Ricci scalar and $\mathcal{L}_m$ corresponds to 
the matter Lagrangian. 

In the metric formalism, the modified Einstein equations (EFEs), obtained by varying this action with respect to the metric takes the form 
 \begin{eqnarray}
 &f'G_{ab}=T^{m}_{ab}+\frac{1}{2}g_{ab}(f-Rf')+\nb_{b}\nb_{a} f'-\gd \nb_{c}\nb^{c} f'\;,\nonumber\\
  \label{efe1}
 \end{eqnarray}
 where  $f\equiv f(R)$, $f'\equiv \frac{{d}f}{{d}R}$ and
$T^{m}_{ab}\equiv \frac{2}{\sqrt{-g}}\frac{\delta(\sqrt{-g}\mathcal{L}_{m})}{\delta \gd}$,
or alternatively
\be
R_{ab}\,=\,\frac{1}{f'}\left[T_{ab}+\frac{1}{2}f g_{ab}-g_{ab}\Box f' +\nabla_a\nabla_b f' \right]\;,
\ee
whose trace is
\be
R=\frac{1}{f'}\left[T+2f-3\Box f'\right]\;.
\ee
Defining the energy momentum tensor of the curvature ``fluid'' (denoted by super or subindex ${\it{R}}$) as
\be
T^{R}_{ab}\equiv\frac{1}{f'}\left[\frac{1}{2}(f-Rf')g_{ab}+\nb_{b}\nb_{a}f'-g_{ab}\nb_{c}\nb^{c}f' \right]\;,
\ee
the field equations (\ref{efe1}) can be written in a more compact form
 \be
 G_{ab}=\tl T^{m}_{ab}+T^{R}_{ab}\;,
 \equiv T_{ab},
 \ee
 where the effective energy momentum tensor of standard matter is given by
 \be
 \tl T^{m}_{ab}\equiv\frac{T^{m}_{ab}}{f'}\;.
 \ee
 Assuming that the energy-momentum conservation of standard matter $T^{m}_{ab}{}^{;b}=0$ holds, this leads us to conclude that $T_{ab}$ is divergence-free, i.e., $T_{ab}{}^{;b}=0$, 
and therefore  $\tl{T}^{m}_{ab}$ and $T^{R}_{ab}$ are not individually conserved \cite{CDT}:
 \be
 \tl T^{m}_{ab}{}^{;b}=\frac{T^{m}_{ab}{}^{;b}}{f'}-\frac{f''}{f^{'2}}T^{m}_{ab}R^{;b}\;,~~~T^{R}_{ab}{}^{;b}=\frac{f''}{f^{'2}}T^{m}_{ab}R^{;b}\;.
 \ee
\subsection{$1+3$ decomposition}
Before proceeding further let us introduce the $1+3$ decomposition. This decomposition will prove to be very useful in the following calculations.
Let us consider the four velocity $u^a$ ($u^c u_c=-1$) and the projection operator defined by
\be
h_{ab}\,=\,g_{ab}+u_a u_b\;,
\label{projector_h}
\ee
which projects into the rest space orthogonal to $u^a$ and satisfies
\be
h_{ab}u^b\,=\,0\,,\;h^{c}_{\;\;a}h^{b}_{\;\;c}\,=\,h^{b}_{\;a}\,,\;\,h^{a}_{\;\;a}=3\;.
\label{properties_h}
\ee
It follows that any spacetime $4$-vector $v_a$ may be covariantly split into a scalar $V$, which is the part of the vector parallel to $u_a$, and a $3$-vector, $V_a$, lying in the sheet orthogonal to $u_a$:
\be
v_a\,=\,-u_a V+V_a\,,\;\;V=v_b u^b\,,\;\; V_a=h^{b}_{\;\;a} v_b\;.
\label{v_decomposition}
\ee
The variation of the velocity with position and time is of interest here and therefore we consider its covariant derivative split into its irreducible parts:
\be
\nabla_a u_b\,=\,D_a u ^b - u_a\dot{u}_b\;,
\label{cov_u}
\ee
then splitting the spatial change of the 4-velocity further into its symmetric and anti-symmetric parts and the symmetric part further into its trace and trace-free part:
\be
\nabla_a u_b\,=\,\sigma_{ab}+\omega_{ab}+\frac{1}{3}\Theta\,h_{ab}-u_a\dot{u}_b\;,
\label{cov_u_bis}
\ee
where $\sigma_{ab}$, $\omega_{ab}$ and $\Theta$ denote the shear tensor, vorticity tensor and expansion scalar respectively.

Applying the previous decomposition to $f(R)$ modified gravity theories in the metric formalism, one gets for general spacetimes:
\begin{eqnarray}
\nabla_a\nabla_b f'\,&=&\,-\,\dot{f}'\left( \frac{1}{3}h_{ab}\Theta+\sigma_{ab}+\omega_{ab}\right)
+u_{b} u_a \ddot{f}' \nonumber\\ &&+\, u_{a}\dot{f}'\dot{u}_b\;,
\label{nablas_f_general}
\end{eqnarray}
and consequently 
\be
\Box f'\,=\, -\Theta\dot{f}'-\ddot{f}'\;, 
\label{box_f_general}
\ee
where terms involving the orthogonally projected derivative have been dropped since our focus is on homogeneous and isotropic spacetimes.
\subsection{The Background FLRW equations}
Considering a flat universe filled with standard matter with energy density $\mu_m$ and pressure $p_m$ with an FLRW metric, the non-trivial field equations obtained from (\ref{lagr f(R)}) lead to the following equations governing the expansion history of the model
\begin{eqnarray}\label{raych}
&&3\dot{H}+3H^2=-\frac{1}{2f'}\left(\mu_m+3p_m+f-f'R+3H f'' \dot{R}\right.\nonumber\\ 
&&~~~~~~~~~~~~~~~~\left.+\,3f'''\dot{R}^2+3f''\ddot{R}\right)
\label{ray}\;,\\
&&3H^2= \frac{1}{f'}\left(\mu_m+\frac{Rf'-f}{2}-3H f'' \dot{R}\right)\;, 
\label{fried}
\end{eqnarray}
i.e., the {\em Raychaudhuri} and {\em Friedmann} equations \cite{Ray,Fried}. Here $H$ is the Hubble parameter, which defines the scale factor $a(t)$ via the standard relation $H=\dot{a}/{a}$ and the Ricci scalar is 
\begin{equation}
R=6\dot{H}+12H^2\;. \label{Riccci}
\end{equation}
The {\em energy conservation equation} for standard matter
\begin{equation}\label{cons:perfect}
\dot{\mu}_m=-3H\mu_m\left(1+w_m\right)
\end{equation}
closes the system, where $w_m$ is its barotropic equation of state.

Note that the Raychaudhuri equation can be obtained from the Friedmann equation,  the energy conservation equation and the definition of the Ricci scalar. Hence, any solution of  the Friedmann equation automatically solves the Raychaudhuri equation. 

In a similar way, we can decompose the  energy momentum tensor of the curvature fluid to obtain the corresponding  thermodynamical quantities (denoted in what follows by a ${\it{R}}$ superscript or subscript). All these quantities, unlike their matter counter-parts, vanish in standard GR, with a FLRW geometry 
\begin{eqnarray}
\mu^{R}&=&T^{R}_{\;ab}u^{a}u^{b}= \frac{1}{f'}\left[\frac{1}{2}(Rf'-f)-\Theta f'' \dot{R} 
\right]\,,\nonumber\\
p^{R}&=&\frac{1}{3}T^{R}_{\;ab}h^{ab}=\frac{1}{f'}\Big[\frac{f-Rf'}{2}+f''\left(\ddot{R}+\frac{2}{3}\Theta\dot{R}\right)
\nonumber\\
&&~~~~~~~~~~~~~~~+\, f'''\dot{R}^{2}
\Big]\,,
\end{eqnarray}
where the anisotropic stress and energy flux (momentum density) vanish in these geometries. With the definition of standard matter and the curvature fluid, one can define a total 
equation of state parameter $\omega_{total}$ as follows:
\be
\omega_{total}\,\,\equiv\,\frac{p_{total}}{\mu_{total}}\,=\,\frac{p_m/f'+p^R}{\mu_m/f'+\mu^R}\;,
\label{omega_total}
\ee
where total density and pressure can be combined, so that
\be
\dot{\mu}_{total}+3H\left(\mu_{total}+p_{total}\right)\,=\,0\;.
\label{conservation_total}
\ee
Let us stress that $\omega_{total}$ does not represent the equation of state of any physical fluid or mixture thereof, 
but should instead be regarded as a mathematical trick that allows us to rewrite the EFEs and the conservation 
equation (\ref{conservation_total}) in a more convenient way. 
\section{Geodesic deviation equation in $f(R)$ gravity}
\label{Section3}
The general GDE takes the form  \cite{Schouten,Synge,Wald} 
\be
\frac{\delta^2 \eta^a}{\delta v^2}\,=\, - R^{a}_{\;\;bcd} V^b V^d \eta^c\;,
\ee
where $\eta_a$ is the deviation vector, $V^a$ is the normalised tangent vector field  and $v$ is an affine parameter.
It is obvious that the contraction of the Riemann tensor with the normalised tangent vector field $V^a$ and the deviation vector $\eta^a$ depend on the tensorial equations provided by the gravitational theory under consideration. In order to make explicit the $f(R)$ dependence in the previous expression, let us consider the usual Weyl tensor definition
\begin{eqnarray}
C_{abcd}\,&=&\,-\frac{1}{2}\left(g_{ac}R_{bd}-g_{ad}R_{bc} +g_{bd}R_{ac} -g_{bc}R_{ad} \right)\nonumber\\
&&+\frac{R}{6}\left(g_{ac}g_{bd}-g_{ad}g_{bc}  \right)+R_{abcd}\,.
\label{Weyl-Riemann}
\end{eqnarray}
For homogeneous and isotropic spacetimes, the Weyl tensor is identically zero and therefore \ref{Weyl-Riemann} when contracted with $V^b \eta^c V^d$ yields
\begin{eqnarray}
R^{a}_{\;bcd}V^b\eta^{c}V^{d}\,&=&\,\frac{1}{2}\left(\eta^aV^bV^dR_{bd}-V^{a}V^{b}\eta^{c}R_{bc}\right.\nonumber\\
&&\left.+\,\epsilon R^{a}_{\,c}\,\eta^c \right)-\frac{R}{6}\eta^a\epsilon\;,
\label{GDE-1}
\end{eqnarray}
with $E=-V_{a} u^{a}$, $\eta_{a} u^a = \eta_a V^a=0$ and $\epsilon = V_a V^a$.
 The terms in (\ref{GDE-1}) can be simplified as follows
\begin{eqnarray}
R^{a}_{bcd}\eta^c        &=& \frac{1}{f'} \left[\eta^a\left(p_m+\frac{f}{2} -\Box f' \right) +\left(\nabla^a\nabla_c f'\right)\eta^c\right]\,, \nonumber\\
R_{bc}V^a V^b \eta^c &=&  \frac{1}{f'} \left[ \left(\nabla_b\nabla_c f' \right)V^aV^b\eta^c   \right]\,, \nonumber\\
R_{bd}V^bV^d\eta^a  &=&  \frac{1}{f'} \Big[\left(\mu_m+p_m\right)E^2+\epsilon\left(p_m +\frac{f}{2} - \Box f' \right) \nonumber\\
 && + V^bV^d\nabla_b\nabla_d f'\Big]\eta^a\;. 
\label{GDE-parts-1}
\end{eqnarray}
When assuming homogeneous and isotropic spacetimes (FLRW), i.e., $\omega_{ab}=0=\sigma_{ab}$ and using (\ref{nablas_f_general}), we get
\begin{eqnarray}
V^bV^d\nabla_b\nabla_d f ' &=&  -\frac{1}{3} \dot{f}'\Theta\left(\epsilon + E^2\right) +E^2\ddot{f}'\,, \nonumber\\
\left(\nabla_b\nabla_c f'\right)V^aV^b\eta^c&=&0\,, \nonumber\\
\left(\nabla^a\nabla_c f'\right)\eta^c &=&-\frac{1}{3}\dot{f}'\Theta\,\eta^a\,.
\label{GDE-parts-2}
\end{eqnarray}
Consequently (\ref{GDE-1}) becomes
\begin{eqnarray}
R^a_{\,bcd}V^bV^d\eta^{c}\,&&=\,\frac{1}{2f'}\left[ \frac{f+\mu_m-2\dot{f}'\Theta}{3} - \Box f'+p_m \right]\eta^a \epsilon\nonumber\\
&&+ \frac{1}{2f'}\left[   \mu_m+p_m-\frac{1}{3}\dot{f}'\Theta+\ddot{f}'  \right]\eta^a E^2\;.
\label{GDE-2}
\end{eqnarray}
Using the fact that 
\begin{eqnarray}
\mu^{R}+p^{R}\,&=&\,\frac{1}{f'}\left[-\frac{1}{3}\dot{f}'\Theta +\ddot{f}'\right]\,,\nonumber\\
\mu^{R}+3p^{R}\,&=&\,\frac{1}{f'}\left[f+\Theta\dot{f}'+3\ddot{f}'\right]-R \,,
\label{rho_plus_P_Curvature}
\end{eqnarray}
we obtain, after some manipulations, the final result for the GDE in $f(R)$ theories within the metric formalism: 
\begin{eqnarray}
R^{a}_{\;bcd}V^{b}V^{d}\eta^{c}\,&=&\,\frac{1}{2}\left(\mu_{total}+p_{total}\right)E^2 \eta^{a}\nonumber\\
&+&\left[ \frac{R}{6}+\frac{1}{6}\left(\mu_{total}+3p_{total}\right) \right]\epsilon\, \eta^a\;.
\label{GDE_fR}
\end{eqnarray}

As expected from the homogeneous and isotropic geometry, the GDE in these type of theories only result in a change in the deviation vector component $\eta^{a}$, i.e., the force 
term is proportional to $\eta^a$ itself and, consequently, according to \cite{GDE-original-1,GDE-original-2} only the magnitude of $\eta$ may change along the geodesic, whereas 
its spatial orientation remains fixed. Note also that the standard GR result is recovered when $f(R)=R$.  If anisotropic geometries are
considered,  a change also in the direction of the deviation vector would result. This analysis will be left to future work.
\section{Null geodesics in $f(R)$ theories}
\label{Section4}
Let us now restrict our investigation to null vector fields, in this case $V^a=k^a$ with $k_a k^a=0$ and consequently $\epsilon=0$. Equation (\ref{GDE_fR}) then reduces to
\begin{eqnarray}
R^{a}_{\;bcd}k^{b}k^{d}\eta^{c}\,=\,\frac{1}{2}\left(\mu_{total}+p_{total}\right)E^2 \eta^{a}\;,
\label{GDE_fR_null}
\end{eqnarray}
which expresses the focusing of all families of past directed geodesics provided that 
\begin{eqnarray}
\left(\mu_{total}+p_{total}\right)\,>\,0
\label{Focusing_null}
\end{eqnarray}
is satisfied. At this stage let us stress that the usual GR result is recovered from (\ref{GDE_fR_null}) and that a cosmological constant term in the gravitational Lagrangian 
with equation of state $p_{\Lambda}=-\rho_{\Lambda}$ does not affect the focusing of null geodesics \cite{Ellis-Elst}. 
Nevertheless, in the realm of modified gravity theories, Eq. (\ref{Focusing_null}) does not need to be satisfied a priori in order to guarantee the viability of a theory or classes of models therein ({\it c.f.} \cite{21} and \cite{Franco2} for thorough discussions on this issue). 

\subsection{Past-directed null geodesics and area distance in $f(R)$ theories}
Let us now consider $V^{a} = k^{a}$, $k_{a}\,k^{a} = 0$, $k^{0} < 0$ and let us study the consequences of equation (\ref{GDE_fR_null}).
Writing $\eta^{a} = \eta\,e^{a}$, $e_{a}\,e^{a}=1$, $0 =
e_{a}\,u^{a} = e_{a}\,k^{a}$, and using a basis which is both parallel propagated and
aligned, i.e., $\delta e^{a}/\delta v = k^{b}\nabla_{b}e^{a}=0$,  one can rearrange (\ref{GDE_fR_null}) as
\begin{eqnarray}
\frac{{\rm d}^2 \eta}{{\rm d}v^2}
\,=\,-\frac{1}{2}\left(\mu_{total}+p_{total}\right)E^2 \eta^{}\;.
\label{GDE_fR_null_bis}
\end{eqnarray}

Provided that $(\mu_{total}+p_{total}) > 0$, all families of past-directed (and future-directed) null geodesics
experience focusing. 
For the pathological case, where the right hand side of  (\ref{GDE_fR_null_bis}) vanishes - in GR this scenario corresponds to a de Sitter universe - the solution 
of (\ref{GDE_fR_null_bis}) becomes $\eta(v) = C_{1}\,v + C_{2}$, equivalent to the case of 
flat (Minkowski) spacetime. 

After some manipulation that involves using expressions (\ref{ray}) and (\ref{fried}), as well as the fact that
\begin{eqnarray}
\frac{{\rm d}^2}{{\rm d}v^2}\,&=&\,
\left(\frac{{\rm d}z}{{\rm d} v}\right)^2
\left[ \frac{{\rm d}^2}{{\rm d} z^2} -\frac{{\rm d}z}{{\rm d}v}\,\frac{{\rm d}^2v}{{\rm d}z^2}\,\frac{{\rm d}}{{\rm d}z} \right]\,,
\nonumber\\
\frac{{\rm d z}}{{\rm d} v}\,&=&\,E_0\, H (1+z)\,, 
\label{derivatives}
\end{eqnarray}
equation (\ref{GDE_fR_null_bis}) in redshift yields
\be
\frac{{\rm d}^2\eta}{{\rm d}z^2}+\frac{(7+3\omega_{total})}{2(1+z)} \frac{{\rm d}\eta}{{\rm d}z}+\frac{3(1+\omega_{total})}{2(1+z)^2}\,\eta\,=\,0\;.
\label{Final_expression}
\ee
It follows that (\ref{Final_expression}) depends only on $\omega_{total}$ as a function of redshift, i.e., as a function of the cosmological evolution. 

Equipped with the previous result, one can infer an expression for the 
observer area distance $r_{0}(z)$:
\begin{eqnarray}
r_{0}(z) := \sqrt{\ \left|\,\frac{{\rm d}A_{0}(z)}{{\rm d}\Omega}\,
\right|\ } = \left| \frac{\left.\eta(z')\,\right|_{z}}{\left.
{\rm d}\eta(z')/{\rm d}\ell\,\right|_{z'=0}}\ \right| \,,
\label{Area distance definition}
\end{eqnarray}
where $A_{0}$ is the area of the object and $\Omega$ the solid angle. We have used the fact that ${\rm d}/{\rm d}\ell = E_{0}^{-1}\,(1+z)^{-1}\,{\rm d}/{\rm d} v =
H\,(1+z)\,{\rm d}/{\rm d}z$ 
and chosen the deviation to be zero at $z=0$. Thus $r_0$ is given by 
\begin{eqnarray}
r_{0}(z)\, =\, \left| \frac{\eta(z)}
{H(0)\,\left.{\rm d}\eta(z')/{\rm d}z'
\,\right|_{z'=0}
}\right|\;.
\label{Area distance final}
\end{eqnarray}
Analytical expression for the observable area distance for GR with no cosmological constant can be found in 
\cite{Ellis-Elst, Matravers-Aziz}, whereas for more general scenarios numerical integration is usually required. 

\section{Dynamical system formalism}
\label{Section5}
Finding solutions of the cosmological field equations (\ref{fried}) - (\ref{cons:perfect}) can in general become a cumbersome issue. We therefore employ a general dynamical systems strategy, following \cite{DS, Dyn}, to significantly simplify the system of equations. For example, rewriting the Friedmann equation at (\ref{fried}) in the following way:
\begin{equation}
H^{2} = \frac{\mu_{m}}{3f'} + \frac{1}{6}R-\frac{1}{6}\frac{f}{f'} - H\frac{f''\dot{R}}{f'}\;,
\end{equation}
leads quite naturally to the definition of the following set of general dimensionless dynamical variables:
\ber
\label{defdsysv}
&& x\equiv\dfrac{\dot{R} f''}{f'H},~~~~~\qquad y\equiv\dfrac{R}{6H^2},~~\qquad \chi\equiv\dfrac{f}{6f' H^2},\nn 
&& \tilde{\Omega}_m\equiv\dfrac{\mu_m}{3f' H^2},\qquad h(z)\equiv\dfrac{H}{H_0}\;.
\eer
Substituting the modified field equations (\ref{fried}) - (\ref{cons:perfect}), for dust, into the redshift derivative of the above variables, leeds to the following set of five first order differential equations
\ber
(1+z)\frac{{\rm d}h}{{\rm d}z}\,&=&\,h\left(2-y\right)\;,\\
(1+z)\frac{{\rm d}x}{{\rm d}z}\,&=&\,x^2+x(y+1)-2y+4\chi-\tilde{\Omega}_m\,,\\
(1+z)\frac{{\rm d}y}{{\rm d}z}\,&=&\,y(2y-xQ-4)\;,\\
(1+z)\frac{{\rm d}\chi}{{\rm d}z}\,&=&\,\chi\left(x+2y-4\right)-x y Q\;,\\
(1+z)\frac{{\rm d}\tilde{\Omega}_m}{{\rm d}z}\,&=&\,\tilde{\Omega}_m\left(x+2y-1\right)\;,
\eer
\noindent with the Friedmann constraint 
\be
\label{const}
1=y-\chi-x+\tilde{\Omega}_{m}\,.
\ee
where the term $Q\equiv\frac{f'}{Rf''}$ specifies the theory under consideration. In order to close the system, $Q$ must be expressed in terms of the dynamical systems variables. 

To solve these equations requires fixing initial conditions for the normalised Hubble parameter $h$ and the deceleration parameter $q$, together with fixing the value of $\Omega_0$ today. In this way we can compute the initial values of  $\{y,\, \chi,\, \tilde{\Omega}_{m} \}$ directly using (\ref{defdsysv}) and $x$ through the constraint (\ref{const}). In general the background evolution will differ from $\Lambda$CDM, leading to a different predictions from the GDE.

In terms of the DS variables introduced in (\ref{defdsysv}), the GDE for models given by (\ref{Model_1}) can be rearranged as follows:
\be
\frac{{\rm d}^2\eta}{{\rm d}z^2}+\frac{4-y(z)}{1+z} \frac{{\rm d}\eta}{{\rm d}z}+\frac{2-y(z)}{(1+z)^2}\,\eta\,=\,0\;,
\label{Final_expression_DS}
\ee
since $\omega_{total}=(1-2y(z))/3$ and we have used (\ref{defdsysv}) and (\ref{const}). In fact (\ref{Final_expression_DS}) remains valid regardless of the $f(R)$ theory under consideration as can be seen by a straightforward calculation.
\section{Results for a class of $f(R)$ theories}
\label{Section6}
To illustrate the results in the previous sections, we consider the following broken power-law form for $f(R)$:
\be
f(R)\,=\, aR-m^2\frac{b\left(\frac{R}{m^2}\right)^n}{1+c\left( \frac{R}{m^2}\right)^n}\,,
\label{Model_1}
\ee
where the constants $a, b$ and $c$ are dimensionless model parameters to be constrained by observations, and $m^{2}$ is related to the square of the Hubble parameter. In what follows we instead use the dimensionless parameter $d \equiv m^{2}/H_{0}^{2}$. 

This form of $f(R)$, proposed by Hu \& Sawicki,  has attracted much interest in the literature as a viable alternate for the gravitational interaction. Its popularity is due to its broken power-law nature. This enables the theory to assume the properties of standard GR in low curvature regimes, as well as mimic the observed late-time accelerated expansion behavior, accurately described by $\Lambda$CDM, in the high curvature regimes. As can be seen by the form of (\ref{Model_1}), there is no explicit cosmological constant term, however, as $R \rightarrow \infty$,  an effective cosmological constant appears, in the limiting case of $b/c \rightarrow 0$, manifesting in a constant valued plateau in the function $f(R)$.  When the initial value of the function (\ref{Model_1}) is chosen such that it lies comfortably on this plateau, an appropriately parameterized HS model mimics the behavior of the $\Lambda$CDM model very well. 

\begin{center}
\begin{table}[ht!]
\begin{tabular}{||c|c|c||}
\hline
\hline
Exponent $n$ & $h_0$ & $q_0$\\
\hline
1 &  0.9405 & -0.2274\\
\hline
1.1&  0.9655 & -0.2986\\
\hline
1.4 & 0.9918 & -0.4224\\
\hline
1.8 & 0.9967 & -0.5051\\
\hline
2 &  0.9983 & -0.5274\\
\hline
\hline
\end{tabular}
\caption{
Present-day values of the Hubble $h_0\equiv H({\rm today})/H_0$ and deceleration ($q_0$) parameters for models of the form (\ref{Model_1}) for different values of exponent $n=$ 1,  1.1, 1.4, 1.8 and 2. 
$H_0$ corresponds to the $\Lambda$CDM Hubble parameter value today. 
%
All the studied models provide $h_0$ values  as well as acceleration today ($q_0<0$) very close to $\Lambda$CDM counterparts. Initial conditions for the cosmological evolution were imposed at $z_{in}=10$ (deep in the Hu-Sawicki plateau) matching the  $H$ and $q$ values as given by $\Lambda$CDM. For illustrative purposes $\Omega_{m}^{0}=0.3$ was considered.
%
%
}
\label{Table_n_12}
\end{table}
\end{center}

\begin{figure*} [htbp] 
\centering
  	\includegraphics[width=0.480\textwidth]{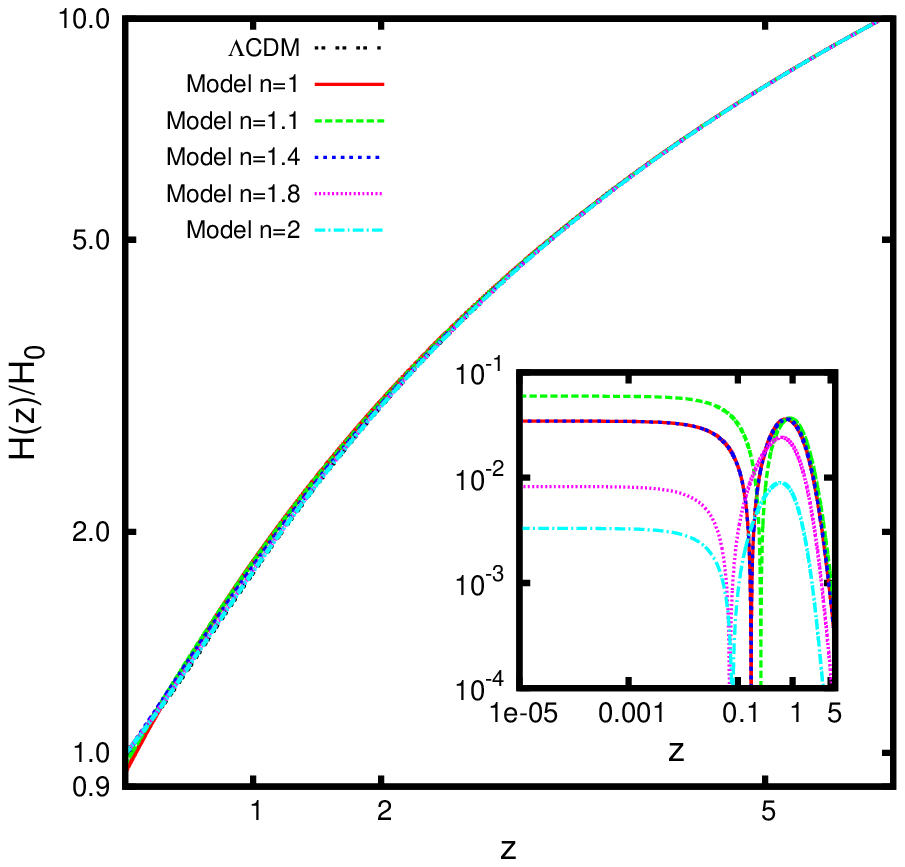}
		\includegraphics[width=0.480\textwidth]{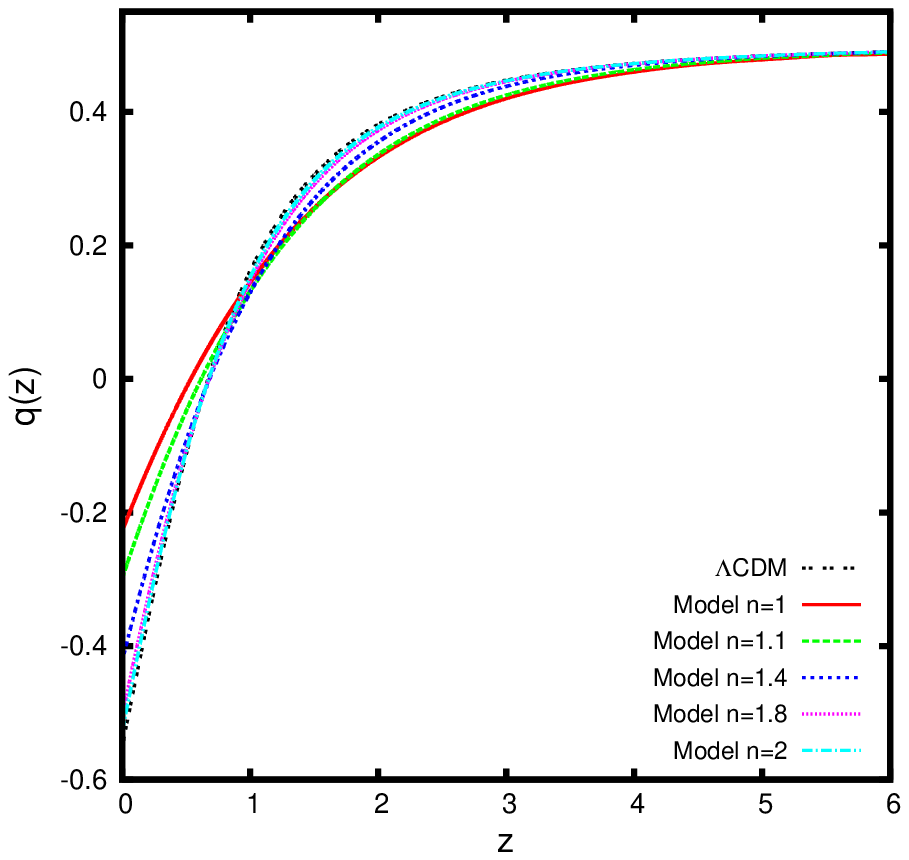}
		\caption{\footnotesize{
Evolution of Hubble parameter  $h=H(z)/H_0$(left panel) and deceleration parameter $q$  (right panel)  as a function of redshift for several exponents $n$. 
$H_0$ holds for the $\Lambda$CDM Hubble parameter value today. For all the models, initial conditions were fixed to match $\Lambda$CDM 
values of $H$ and $q$ at  initial redshift  $z_{in}=10$, i.e., deep in the plateau for this class of $f(R)$ theories.
All the models provide values of $h_0 \approx 1$ as well as acceleration today ($q_0<0$). Explicit values are provided in Table \ref{Table_n_12}.
For illustrative purposes we considered $\Omega_m^0=0.3$ and no radiation.
}}
  \label{Figure_0}
\end{figure*}

\begin{figure*} [htbp] 
\centering
  	\includegraphics[width=0.480\textwidth]{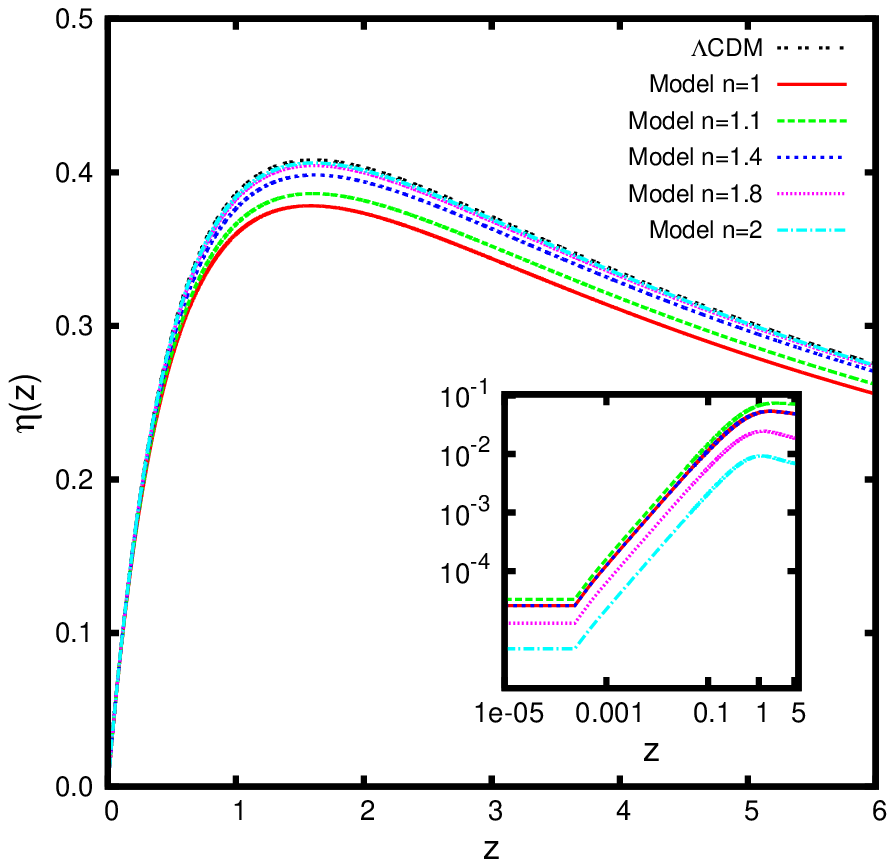}
		\includegraphics[width=0.480\textwidth]{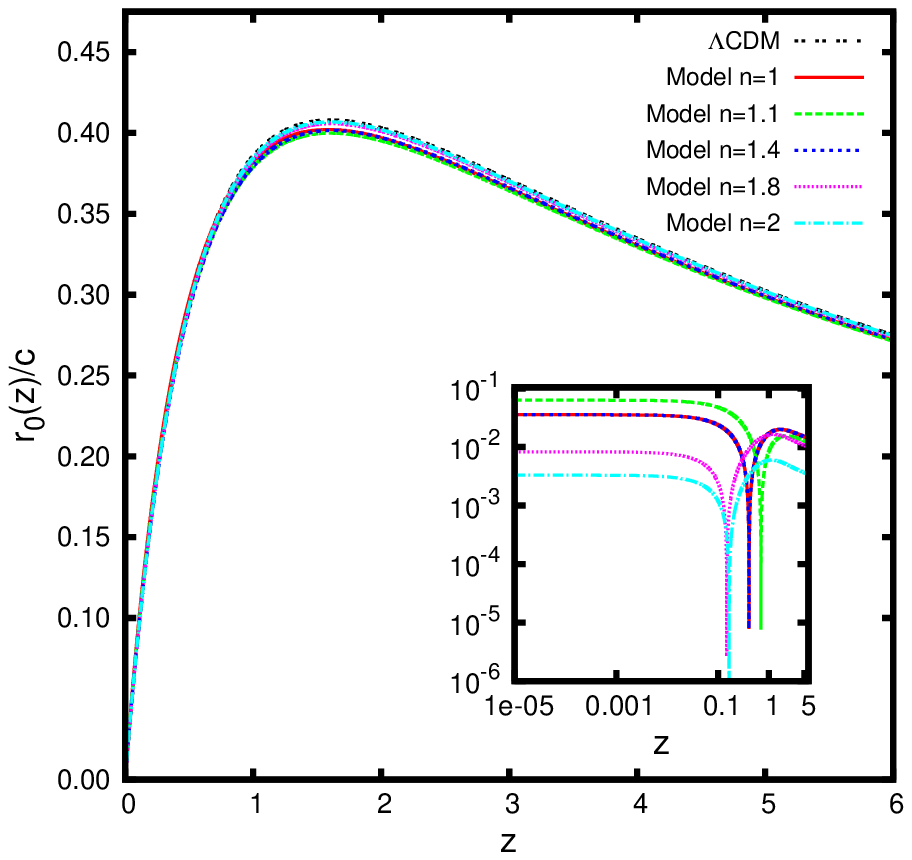}
		\caption{\footnotesize{
		Geodesic deviation $\eta$ (left panel) and area distance (right panel) as functions of redshift. Initial conditions $\eta(z=0)=1$ and $\eta'(z=0)=0$ were imposed in 
		equation (\ref{Final_expression}). The depicted redshift interval was $[0, 6]$. Deviation evolutions are very similar to $\Lambda$CDM. This fact is inherited by 
		the $r_0(z)$ evolution. The inner panels represent relative deviation with respect to $\Lambda$CDM ($\Omega_m^0=0.3$ and no radiation) predicted evolutions 
		for $\eta$ and $r_0$ respectively.
}}
  \label{Figure_1}
\end{figure*}

By specifying the model parameters $\{a, b, c, d, n\}$ and initial values for the dimensionless Hubble parameter, $h_{in}$, and deceleration parameter, $q_{in}$ at an initial redshift $z_{in}$, we can fix the initial values of the dynamical variables $y, \chi$ and $\tilde{\Omega}_{m}$. The constraint equation (\ref{const}) can be used to initialize $x$. 
In order for the HS model to mimic $\Lambda$CDM as closely as possible, the values of $h_{in}$ and $q_{in}$ are set to their corresponding $\Lambda$CDM values, at the chosen initial redshift and study models of the type (\ref{Model_1}) with the fixing of $a=b=1$, $c=1/19$ and $d=6c(1-\Omega_m^0)$ with $\Omega_m^0=0.3$ for illustrative purposes\footnote{The constraint $c=6d(1-\Omega_m^0)$ was considered in order to guarantee that  $\lim_{R \to \infty} f(R) = R- 2\Lambda$ and therefore GR is recovered at the early stages of the Universe.}  and varying the exponent $n$ in the interval $[1,\,2]$. The studied values were $n=$ 1,  1.1, 1.4, 1.8 and 2.
Fig. \ref{Figure_0}  depicts the evolution of Hubble parameter  and deceleration parameter of the aforementioned models.
In Fig. \ref{Figure_1} we have depicted the evolution of the deviation $\eta$ as given by 
$\Lambda$CDM and several $f(R)$ models of the type (\ref{Model_1}) and whose parameters as well as cosmological evolutions are summarized in Table \ref{Table_n_12}. The right panel of Fig. \ref{Figure_1} then showcases the area distance evolution as well as its deviation from $\Lambda$CDM evolution.


For all the studied models, the null geodesic deviation is very similar to the $\Lambda$CDM counterpart having assumed the same standard matter abundance today in all the models. The relative deviation with respect to the $\Lambda$CDM geodesic deviation remains almost indistinguishable (order $10^{-5}$) for very low redshifts and starts to deviate for redshifts $z\approx0.5$ with a relative deviation of order $1\%$. The $\Lambda$CDM evolution seems to constitute an upper bound for the geodesic deviations in all the studied $f(R)$ models, with the relative difference smaller for bigger values of the exponent $n$. Thus $n=2$ model provides the closest geodesic deviation evolution to the concordance model. With respect to the area distance the evolutions resemble with high accuracy that of $\Lambda$CDM, although the latter evolution does not constitute a bound for this quantity. Again $n=2$ provides an area distance evolution almost indistinguishable 
from $\Lambda$CDM in the studied redshift range with a relative deviation smaller than $10^{-2}$. 

As a next step, the equations for the area distance (\ref{Area distance definition}) and 
(\ref{Area distance final}) can be used to constrain these models using several observational probes. For instance, the use of compact radio 
sources as cosmic rulers \cite{gurvits}, the angular size-redshift relation derived from the Sunyaev-Zel'dovich effect - X-ray technique \cite{clusters}. By applying the relation between luminosity distances and area distances it is also possible to extend our studies with type Ia supernovae data \cite{sne} for both homogeneous and statistically homogeneous cases \cite{dr}. 

\section{Conclusions} 
\label{Section7}

In this paper we have presented a complete analysis of the geodesic deviation equation in the metric formalism of $f(R)$ theories. We used a $1+3$ decomposition which enabled us to simplify the intermediate calculations and determine that the new geometrical contributions contribute to the deviation for both null and timelike geodesics. Equation (\ref{GDE_fR})  encapsulates the general result for isotropic and homogeneous geometries. 

We proved that the extra terms introduced by these theories, as well as the standard matter content impact on
the evolution of the geodesic deviation, as is clearly represented in the aforementioned equation. The well-known fact that modified gravity theories do not need to accomplish the standard energy conditions, which standard fluids do \cite{Franco2}, may lead the geodesic deviation equation to exhibit a model-dependent behavior that may serve to constrain  the viability of classes of  models in such theories.

We have illustrated our results for a class of fourth order gravity theories, the so-called Hu-Sawicki $f(R)$ models, which can be considered as a natural extension to the Einstein-Hilbert Lagrangian, able to recover the General Relativity predictions at high curvatures and to provide late-time acceleration, while also satisfying weak field constraints.
First we solved the background equations for different values of the exponents $n$ after having fixed the remaining parameters, where the initial conditions were imposed in the matter dominated epoch, with Hubble and  deceleration parameters matching their $\Lambda$CDM counterparts. 
Let us remind that the initial conditions are fixed well deep in the $f(R)$ Hu-Sawicki model plateau which appears for large curvatures. Therefore for such initial redshift 
the models effectively behave as $\Lambda$CDM once the $f(R)$ model parameters are chosen adequately.
We then used the cosmological background to 
study the evolution of the deviation for null geodesics as well 
as present numerical results for the area distance formula.
For all the cases considered the results are similar to $\Lambda$CDM, which means that they remain phenomenologically 
viable and can be tested with observational data.

The analysis performed in this communication is easily extensible to other $f(R)$ models and modified gravity theories. Work in this direction is in progress in order to apply our results to the most competitive fourth-order gravity as well as scenarios combining gravity theories beyond General  Relativity in non-FLRW spacetimes.

%
\vspace{0.3cm}
{\bf Acknowledgments:} 
A.d.l.C.D. acknowledges financial support from Marie Curie - Beatriu de Pin\'os contract BP-B00195 Generalitat de Catalunya, ACGC fellowship University of Cape Town and 
 MINECO (Spain) projects numbers FIS2011-23000, FPA2011-27853-C02-01 and Consolider-Ingenio MULTIDARK CSD2009-00064.
 A.d.l.C.D. thanks the KITPC, Chinese Academy of Sciences for its hospitality during the final stages of this work.
P.K.S.D. thanks the NRF for financial support. 
V.C.B. is supported by CNPq-Brazil through a fellowship within the program Science without Borders.

\end{document}